\documentclass[aps,showpacs]{revtex4-1}
\usepackage{amsthm}
\usepackage{amsmath,amsfonts,amssymb}
\usepackage{graphicx}

\begin{document}

\title{Bound states in a hyperbolic asymmetric double-well}

\author{R. R. Hartmann}
\email[]{richard.hartmann@dlsu.edu.ph}
\affiliation{
Physics Department,
De La Salle University,
2401 Taft Avenue,
Manila,
Philippines
}

%\date{15 June 2013}
\date{22 November 2013}

\begin{abstract}
We report a new class of hyperbolic asymmetric double-well whose bound
state wavefunctions can be expressed in terms of confluent Heun functions.
An analytic procedure is used to obtain the energy eigenvalues and the criterion for the potential to support bound states is discussed.
\end{abstract}


\maketitle

\section{Introduction}

Many problems in physics from astronomy \cite{Shod_PRL} through to relativity \cite{Hartmann_PRB_10} can be reduced down to Heun's
equation (see \cite{Hortacsu_arXiv} and references therein for a general review). Confluent forms of the Heun
differential equation are obtained when two or more of the regular singularities coalesce to form an irregular one. Many potentials for the Schr{\"o}dinger equation have been shown to transform into the Heun equation and its confluent forms \cite{Hartmann_PRB_10,Chin,Charles_JMP_13,Hartmann_arXiv}.

The asymmetric double-well has been studied across many fields of physics from heterostructures \cite{Fujiwara_PRL_97} and  Bose-Einstein condensates in a double trap \cite{Schumm_Nature_05} to superconducting circuits involving tuneable asymmetric double-wells \cite{Simmonds_PRL_04,Cooper_PRL_04,Johnson_PRL_05}, the latter have attracted a great deal of attention due to their potential use as quantum bits. The asymmetric double-well eigenvalue problem has also been studied via supersymmetry techniques \cite{Gangopadhyaya_PRA}.

We study theoretically a new class of asymmetric double-well, composed of hyperbolic functions with three fitting parameters. It shall be shown that such a potential allows one to reduce the one dimensional Schr{\"o}dinger equation down to the confluent Heun equation. An analytic procedure is used to obtain the energy eigenvalues namely, the eigenvalues are found by calculating the zeros of the Wronskian formed by two Frobenius solutions, each one expanded about the confluent Heun equation's different regular singularities.

\section{Bound states in a hyperbolic asymmetric double-well}

The time independent Schr{\"o}dinger equation reads
\begin{equation}
-\frac{d\Psi}{dx^{2}}+V\left(x\right)\Psi=\varepsilon\Psi.
\label{eq:Se}
\end{equation}
Here on in all energies are measured in units of $2m/\hbar^{2}$ and the model potential under consideration, $V\left(x\right)$, is given by
\begin{equation}
V\left(x\right)=\left\{ -V_{1}\left[1+\tanh^{2}\left(\frac{x}{L}\right)\right]+\left[V_{2}-V_{3}\tanh\left(\frac{x}{L}\right)\right]\right\} \left[1-\tanh^{2}\left(\frac{x}{L}\right)\right],
\label{eq:Potential}
\end{equation}
where the parameters $V_{1}$, $V_{2}$, $V_{3}$ and $L$ characterize the potential strength and width.
For the case of $V_{1}=V_{3}=0$, the potential becomes
the P{\"o}schl-Teller potential which can be
solved exactly, and the wavefunctions are given in terms of Legendre
functions \cite{Poschl}. For the case of $V_{3}=0$, the potential
becomes the Manning potential \cite{Manning_CP_35} which can be used to describe
a harmonic double-well. Substituting equation~(\ref{eq:Potential}) into
equation~(\ref{eq:Se}) and making the change of variable $\xi=\left[1+\tanh\left(z\right)\right]/2$ allows equation~(\ref{eq:Se}) to be written as
\begin{equation}
4\xi^{2}\left(1-\xi\right)^{2}\frac{\partial^{2}\Psi}{\partial\xi^{2}}+4\xi\left(1-\xi\right)\left(1-2\xi\right)\frac{\partial\Psi}{\partial\xi}-4\xi\left(1-\xi\right)\left[-4w_{1}\xi^{2}+2\left(2w_{1}-w_{3}\right)\xi-2w_{1}+w_{2}+w_{3}\right]\Psi=E\Psi
\label{eq:SE_1}
\end{equation}
where we use the dimensionless variable $z=x/L$ with $w_{i}=L^{2}V_{i}$,
$i=1,\,2,\,3$ and $E=-L^{2}\varepsilon$. Using the transformation
$\Psi=\xi^{q}\left(1-\xi\right)^{r}e^{s\xi}H$ allows equation~(\ref{eq:SE_1})
to be reduced to
\begin{equation}
\frac{\partial^{2}H}{\partial\xi^{2}}+\left[\alpha+\frac{1+\beta}{\xi}+\frac{1+\gamma}{\xi-1}\right]\frac{\partial H}{\partial\xi}+\frac{\mu\xi+\nu}{\xi\left(\xi-1\right)}H=0\label{eq:Heun_1}
\end{equation}
where
\begin{eqnarray}
\nonumber
\mu &=&\delta+\alpha\left(\frac{\beta+\gamma+2}{2}\right)\\
\nonumber
\nu &=&\eta+\frac{\beta}{2}+\frac{\left(\gamma-\alpha\right)\left(\beta+1\right)}{2}\\
\nonumber
\end{eqnarray}
and $q$ can take upon the values $\pm\frac{1}{2}\sqrt{E}$ while
$s$ can take upon the values $\pm2\sqrt{w_{1}}$. For the case
of $r=q$
\begin{equation}
%\fl
\alpha=2s\qquad\beta=\gamma=2q\qquad\delta=-2w_{3}\qquad\eta=-2w_{1}+w_{2}+w_{3}+\frac{E}{2}
\nonumber
\end{equation}
while for $r=-q$
\begin{equation}
%\fl
\alpha=2s\qquad\beta=2q\qquad\gamma=-2q\qquad\delta=-2w_{3}\qquad\eta=-2w_{1}+w_{2}+w_{3}+\frac{E}{2}.
\nonumber
\end{equation}
equation~(\ref{eq:Heun_1}) has regular singularities at $\xi=0$ and
$1$, and an irregular singularity of rank $1$ at $\xi=\infty$.
$H$ is the confluent Heun function \cite{Heun,Ronveaux} given by the expression
\begin{equation}
H=H\left(\alpha,\,\beta,\,\gamma,\,\delta,\,\eta,\,\xi\right)=\sum_{n=0}^{\infty}c_{n}\xi^{n}
\nonumber
\end{equation}
where the coefficients $c_{n}$ obey the three term recurrence relation
\begin{equation}
A_{n}c_{n}=B_{n}c_{n-1}+C_{n}c_{n-2}
\nonumber
\end{equation}
with the initial conditions $c_{n-1}=0$ and $c_{n}=1$ where
\begin{eqnarray}
\nonumber
A_{n} &=& 1+\frac{\beta}{n}\\
\nonumber
B_{n} &=& 1+\frac{1}{n}\left(\beta+\gamma-\alpha-1\right)+\frac{1}{n^{2}}
\left[\eta-\frac{1}{2}\left(\beta+\gamma-\alpha\right)-\frac{1}{2}\beta\left(\alpha-\gamma\right)\right]\\
\nonumber
C_{n} &=& \frac{\alpha}{n^{2}}\left(\frac{\delta}{\alpha}+\frac{\beta+\gamma}{2}+n-1\right).\\
\nonumber
\end{eqnarray}
The solutions to equation~(\ref{eq:SE_1}) are therefore given by
\begin{equation}
%\fl
\Psi_{1}=D_{1}\xi^{q}\left(1-\xi\right)^{q}e^{s\xi}H\left(2s,\,2q,\,2q,\,-2w_{3},\,-2w_{1}+w_{2}+w_{3}+\frac{E}{2},\,\xi\right)\label{eq:Psi_1}
\end{equation}
\begin{equation}
%\fl
\Psi_{2}=D_{2}\xi^{q}\left(1-\xi\right)^{-q}e^{s\xi}H\left(2s,\,2q,\,-2q,\,-2w_{3},\,-2w_{1}+w_{2}+w_{3}+\frac{E}{2},\,\xi\right)\label{eq:Psi_2}
\end{equation}
where $D_{1}$ and $D_{2}$ are constants.

Under certain conditions the confluent Heun function can be reduced
to a finite polynomial of order $N$. This occurs when two criteria are met \cite{Ronveaux}:
\begin{equation}
\delta=-\alpha\left(N+1+\frac{\beta+\gamma}{2}\right)\label{eq:Term_cond_1}
\end{equation}
and
\begin{equation}
c_{N+1}=0\label{eq:Term_cond_2}
\end{equation}
where $N$ is a positive integer. Analytic expressions for the energy eigenvalues can be obtained from
equation~(\ref{eq:Term_cond_1}) with the caveat that the potential parameters
$w_{1}$,~$w_{2}$ and $w_{3}$ are interrelated such that the second
termination condition, equation (\ref{eq:Term_cond_2}) is satisfied.
In this instance, the potential belongs to a class of quantum models which are quasi-exactly solvable \cite{Turbiner_JETP_88,Ushveridze_94,Bender_JPA_98,Charles_JMP_13,Hartmann_arXiv}, where only some of the eigenfunctions and eigenvalues are found explicitly.
This method has been applied to calculate the energy levels in various symmetric hyperbolic double-wells \cite{Chin,Charles_JMP_13}. To determine the bound state energies for a potential described by an arbitrary set of potential parameters we require that the wavefunction
vanishes at infinity i.e. $\Psi\left(\xi=0\right)=\Psi\left(\xi=1\right)=0$.
However, the function $H\left(\alpha,\,\beta,\,\gamma,\,\delta,\,\eta,\,\xi\right)$
is only analytic within the disk $\left|\xi\right|<1$. An analytic
continuation of the confluent Heun function can be obtained by expanding the
solution about the second regular singularity $\xi=1$. By relating
the two Frobenius solutions one can obtain the bound state energies
for arbitrary values of the parameters. The second set of solutions can be
constructed by making the change of variable $\xi'=1-\xi$, in this
instance equation~(\ref{eq:Heun_1}) becomes
\begin{equation}
\frac{\partial^{2} H}{\partial\xi'^{2}}
+\left[\widetilde{\alpha}+\frac{1+\widetilde{\beta}}{\xi'}
+\frac{1+\widetilde{\gamma}}{\xi'-1}\right]\frac{\partial H}{\partial\xi'}
+\frac{\widetilde{\mu}\xi'+\widetilde{\nu}}{\xi'\left(\xi'-1\right)} H
=0.
\label{eq:Heun_2}
\end{equation}
For the case of $r=q$
\begin{equation}
%\fl
\tilde{\alpha}=-2s\qquad
\tilde{\beta}=\tilde{\gamma}=2q\qquad
\tilde{\delta}=2w_{3}\qquad\tilde{\eta}=-2w_{1}+w_{2}-w_{3}+2q^{2}
\nonumber
\end{equation}
while for $r=-q$
\begin{equation}
%\fl
\tilde{\alpha}=-2s\qquad\tilde{\beta}=-2q\qquad\tilde{\gamma}=
2q\qquad\tilde{\delta}=2w_{3}\qquad\tilde{\eta}=-2w_{1}+w_{2}-w_{3}+2q^{2}.
\nonumber
\end{equation}
The solutions to equation~(\ref{eq:SE_1}) are therefore given by
\begin{equation}
%\fl
\Psi_{3}=D_{3}\xi^{q}\left(1-\xi\right)^{q}e^{s\xi}H\left(-2s,\,2q,\,2q,\,2w_{3},\,-2w_{1}+w_{2}-w_{3}+2q^{2},\,1-\xi\right)\label{eq:Psi_3}
\end{equation}
\begin{equation}
%\fl
\Psi_{4}=D_{4}\xi^{q}\left(1-\xi\right)^{-q}e^{s\xi}H\left(-2s,\,-2q,\,2q,\,2w_{3},\,-2w_{1}+w_{2}-w_{3}+2q^{2},\,1-\xi\right)\label{eq:Psi_4}
\end{equation}
where $D_{3}$ and $D_{4}$ are constants.

\begin{figure}[h]
\begin{center}
\includegraphics*[height=120mm,angle=270]{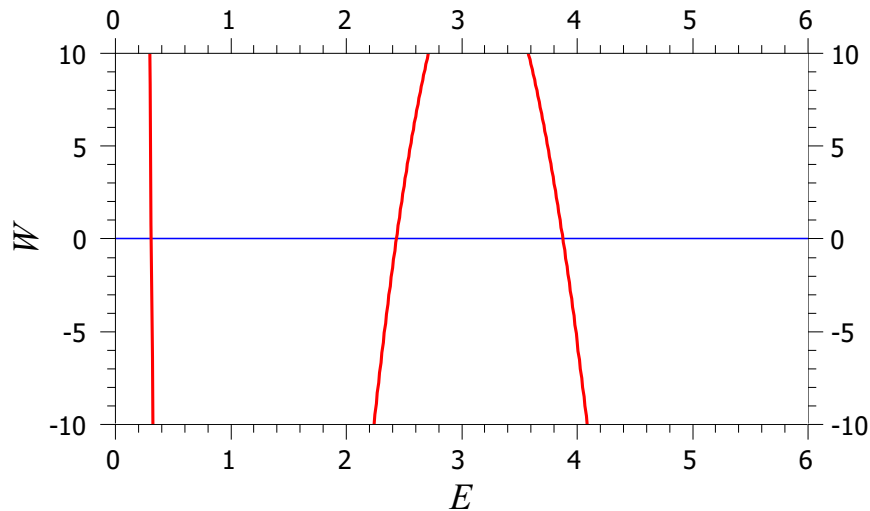}
\end{center}
\caption{
The Wronskian, equation~(\ref{eq:Wronskian}), shown in red, for the hyperbolic asymmetric double-well
as a function of $E$ with $z_{1}=0.2$ for the case of the potential
parameters $w_{1}=15$,~$w_{2}=12$ and $w_{3}=1$. The energy eigenvalues are found when the function $W\left(\Psi_{1},\,\Psi_{3}\right)$ is zero this occurs at $E=0.311$,~$2.434$ and $3.875$. The $W=0$ (in blue) is shown as a guide to the eye.
}
\label{fig:Wronskian}
\end{figure}

\begin{figure}[h]
\begin{center}
\includegraphics*[height=120mm,angle=270]{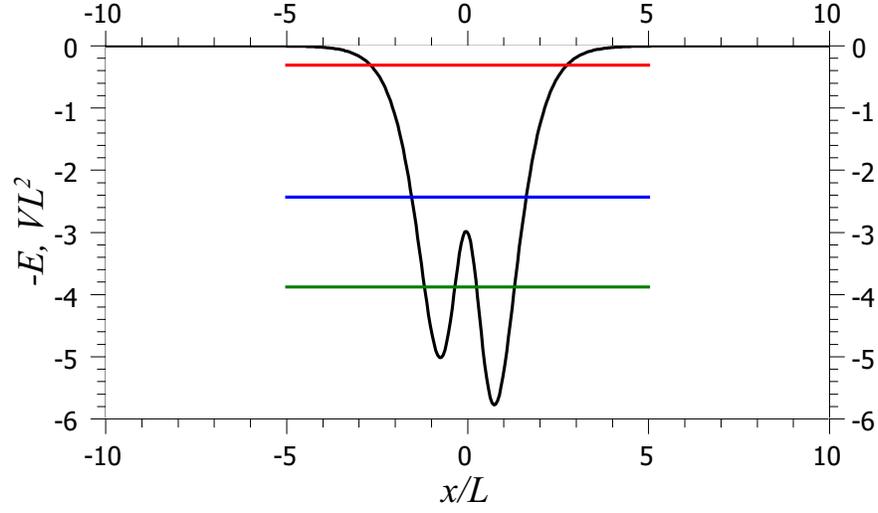}
\end{center}
\caption{
Schematic diagram of the eigenvalue spectrum for the hyperbolic asymmetric double-well described by the parameters
$w_{1}=15$,~$w_{2}=12$ and $w_{3}=1$, in this instance there are three eigenvalues: $E=0.311$,~$2.434$ and $3.875$ and the
potential profile is shown in the same scale.
}
\label{fig:Energy}
\end{figure}

\begin{figure}[h]
\begin{center}
\includegraphics*[height=120mm,angle=270]{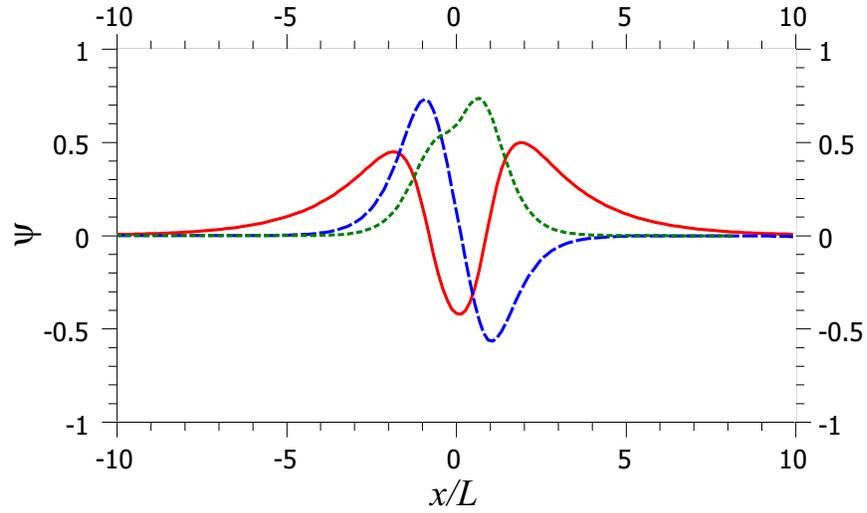}
\end{center}
\caption{
The wavefunctions of the bound states contained within the hyperbolic asymmetric double-well described by the parameters
$w_{1}=15$,~$w_{2}=12$ and $w_{3}=1$. The solid (red), long-dashed (blue) and short-dashed (green) lines correspond to the $E=0.311$,~$2.434$ and $3.875$ eigenvalues respectively. %The potential profile (black line) is shown as guide to the eye.
}
\label{fig:Wavefunctions}
\end{figure}

For $\Psi_{1}$ and $\Psi_{3}$ to be non-divergent functions we require
that $q=\frac{1}{2}\sqrt{E}$. $\Psi_{2}$ ($\Psi_{4}$) requires
$q=-\frac{1}{2}\sqrt{E}$ ($q=\frac{1}{2}\sqrt{E}$) and that the
confluent Heun function is reduced to a confluent Heun polynomial
of the order $N$, where $N>\left|q\right|$ . Equation (\ref{eq:Psi_1}) and equation (\ref{eq:Psi_3})
alone are sufficient to determine the eigenvalue spectrum. The solution
about $\xi=0$ is convergent for $\left|\xi\right|<1$ where as the
solution about $\xi=1$ is convergent for $\left|\xi'\right|<1$.
Therefore providing $z=z_{1}$ lies in both the analytic domains of $\Psi_{1}$
and $\Psi_{3}$ one can write
\begin{equation}
\Psi_{1}\left(z_{1}\right)=\Psi_{3}\left(z_{1}\right)\label{eq:Con_1}
\end{equation}
For the function to be continuous we also require
\begin{equation}
\left.\frac{\partial\Psi_{1}}{\partial z}\right|_{z=z_{1}}=\left.\frac{\partial\Psi_{3}}{\partial z}\right|_{z=z_{1}}.
\label{eq:Con_2}
\end{equation}
Combining equation~(\ref{eq:Con_1}) and equation~(\ref{eq:Con_2}) yields
\begin{equation}
W\left(\Psi_{1},\,\Psi_{3}\right)\left(z_{1}\right)=
\left|
\begin{array}{ccc}
\Psi_{1} & \Psi_{3}\\
\frac{\partial\Psi_{1}}{\partial z} & \frac{\partial\Psi_{3}}{\partial z}
\end{array}
\right|=0.
\label{eq:Wronskian}
\end{equation}
%\begin{equation}
%W\left(\Psi_{1},\,\Psi_{3}\right)\left(z_{1}\right)=
%\begin{vmatrix}\Psi_{1} & \Psi_{3}\\
%\frac{\partial\Psi_{1}}{\partial z} & \frac{\partial\Psi_{3}}{\partial z}
%\end{vmatrix}=0.
%\label{eq:Wronskian}
%\end{equation}
The energy eigenvalues are therefore obtained by finding the zeros
of equation~(\ref{eq:Wronskian}). The Wronskian is comprised of two confluent Heun functions each
corresponding to the Frobenius solutions about the two regular
singularities. In figure \ref{fig:Wronskian}, we plot $W\left(\Psi_{1},\,\Psi_{3}\right)$ with $z_{1}=0.2$ for the potential parameters $w_{1}=15$,~$w_{2}=12$ and $w_{3}=1$ as a function of $E$. The potential is found to contain three bound states at energies $E=0.311$,~$2.434$ and $3.875$. The corresponding energy level diagram and normalized wavefunctions are shown in figure \ref{fig:Energy} and \ref{fig:Wavefunctions} respectively.

\section{Discussion}
\begin{figure}[h]
\begin{center}
\includegraphics*[height=80mm,angle=270]{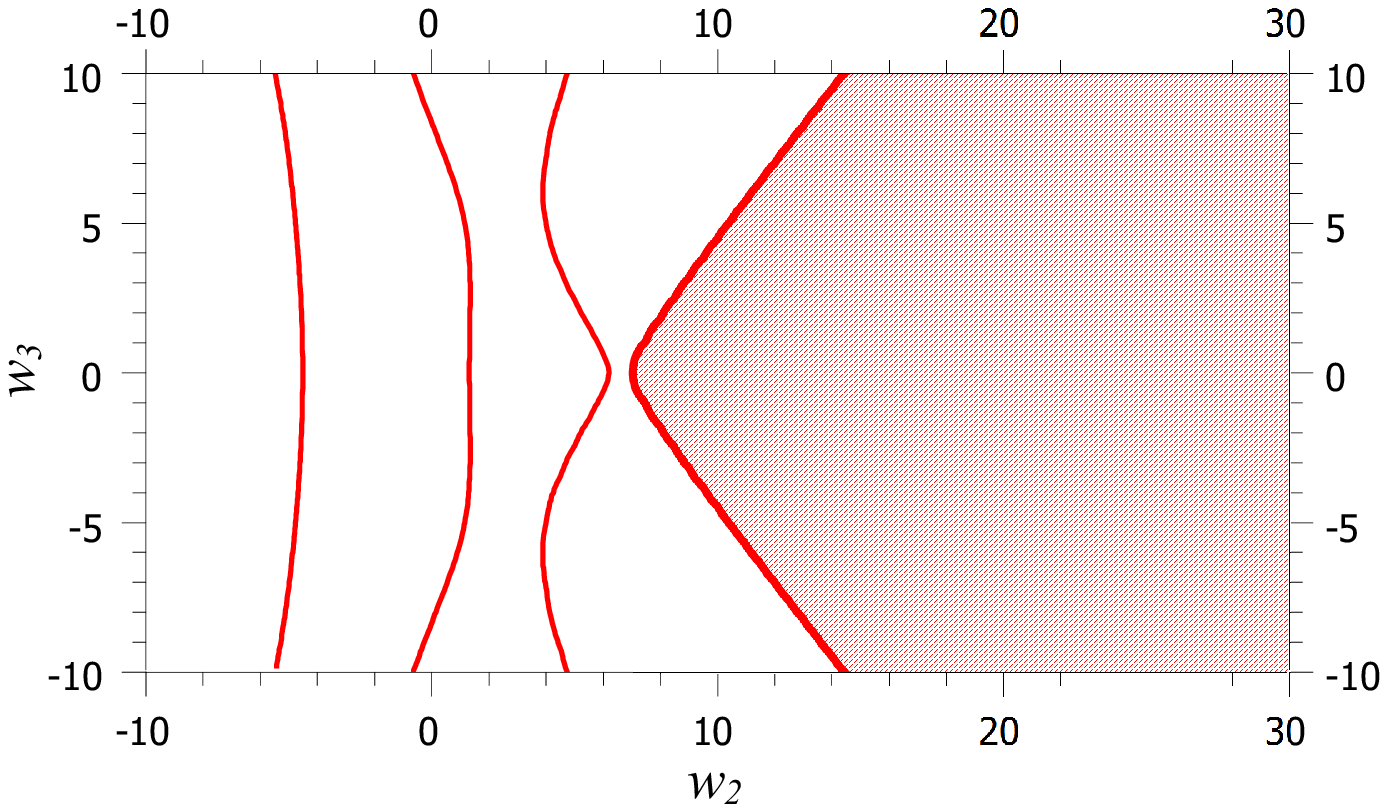}
\includegraphics*[height=80mm,angle=270]{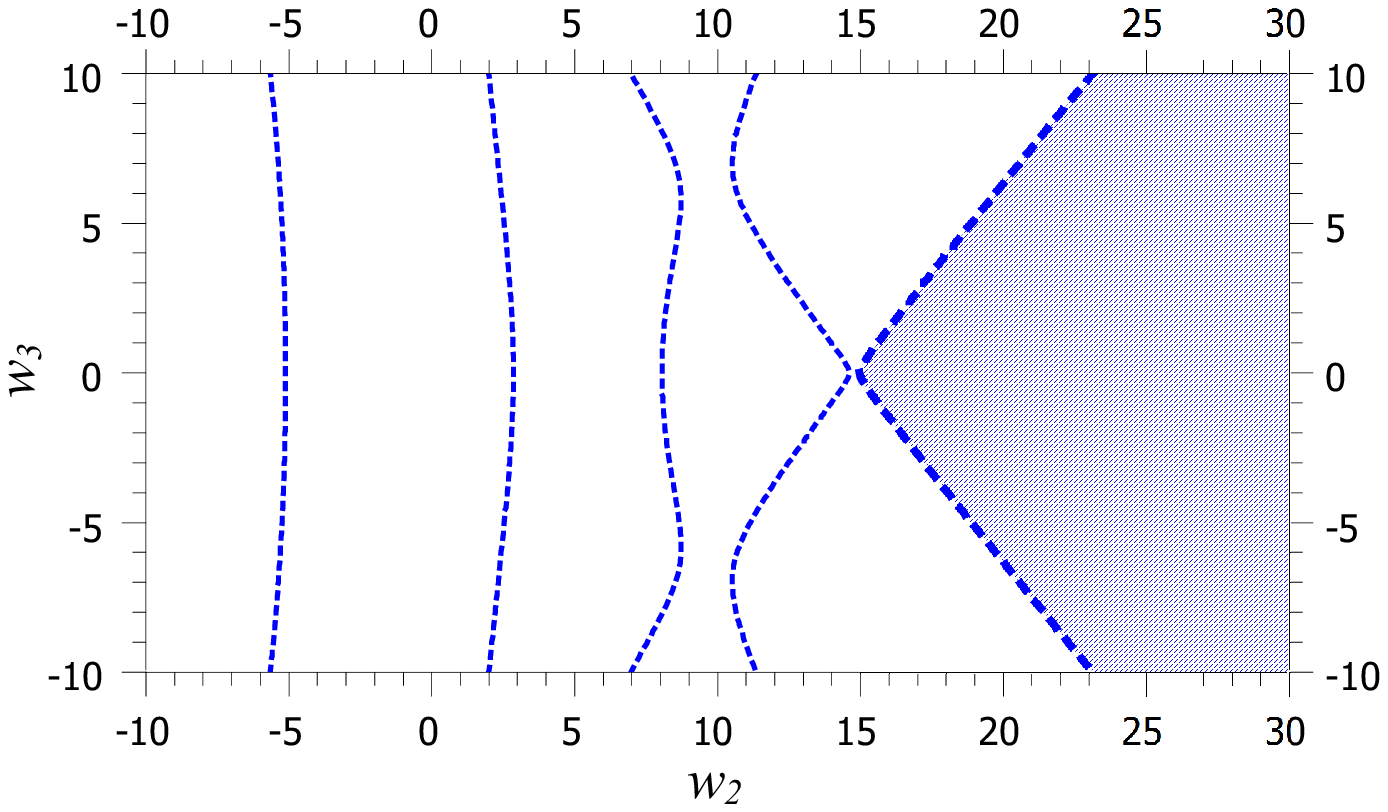}\\
\includegraphics*[height=80mm,angle=270]{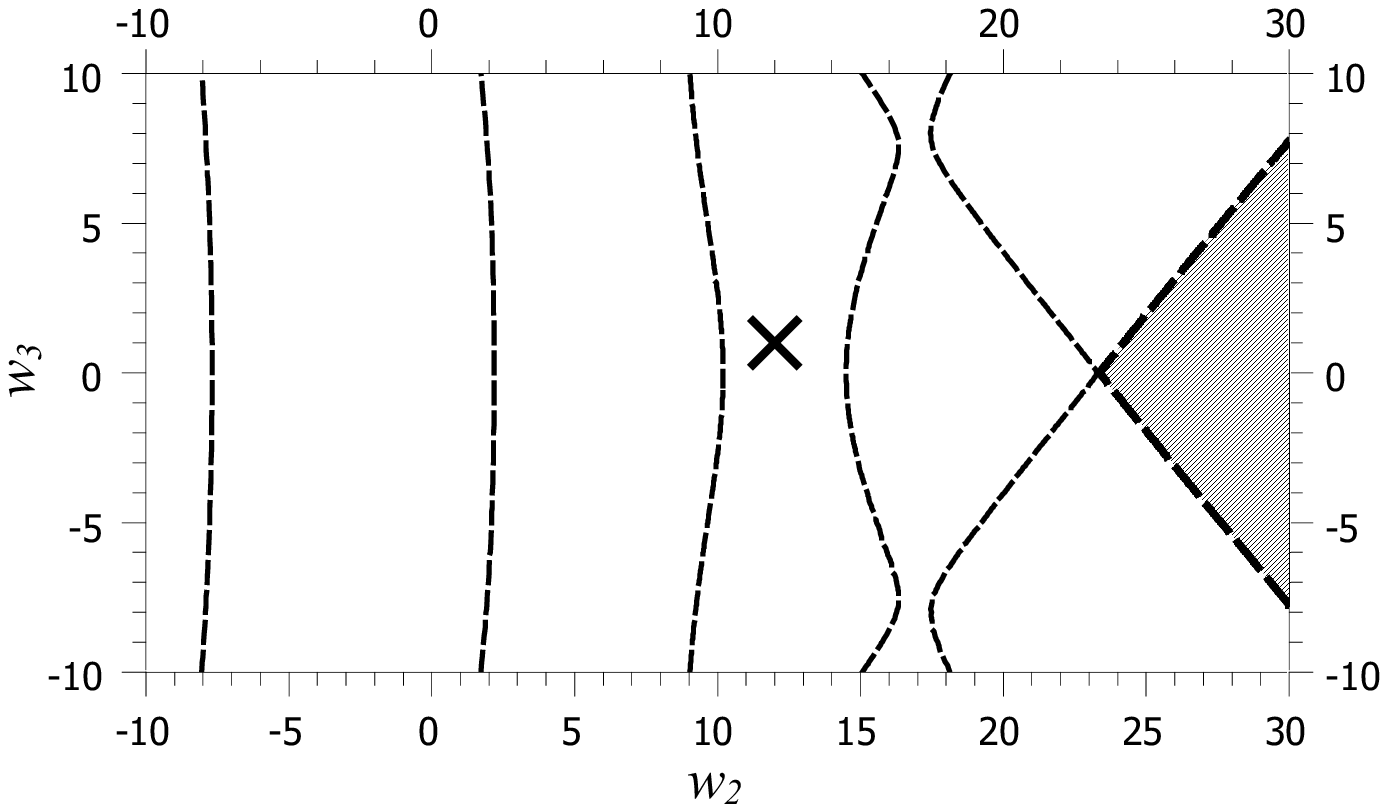}
\includegraphics*[height=80mm,angle=270]{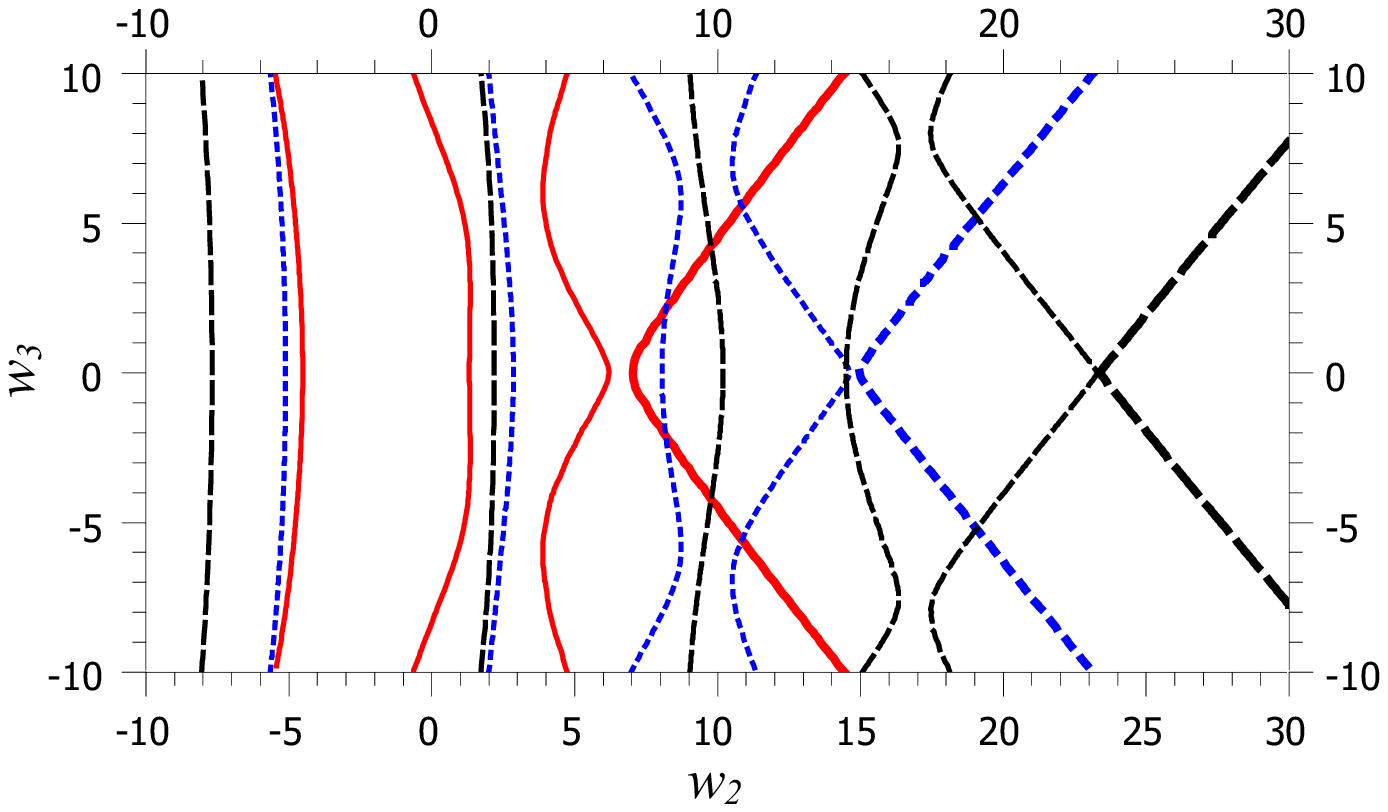}
\end{center}
\caption{
%The combinations of $w_{2}$ and $w_{3}$ which gives rise to potentials which do not contain a bound state are shown for the case of $w_{1}=5$,~$10$ and $15$ which are shaded in red, blue and black respectively.
The combinations of $w_{2}$ and $w_{3}$ which gives rise to potentials which do not contain a bound state are shown for the case of $w_{1}=5$ (shaded in red),~$w_{1}=10$ (shaded in blue) and $w_{1}=15$ (shaded in black). The threshold values of $w_{i}$ for which new bound states emerge from the continuum are also shown by the solid (red), short-dashed (blue) and long-dashed (black) lines which correspond to $w_{1}=5$,~$10$ and $15$ respectively. The critical values of $w_{i}$ which assures that the potential contains a bound state, i.e. the emergence of the first bound state from the continuum, are denoted by the thick lines. The cross corresponds to the potential defined by the combination $w_{1}=15$,~$w_{2}=12$ and $w_{3}=1$.
}
\label{fig:critical}
\end{figure}

The number of bound states contained within the potential is a function of potential strength and width. However, it should be noted that not all combinations of $w_{i}$ result in a potential that can support bound states (see figure~\ref{fig:critical}). Apart from the trivial case wherein combinations of $w_{i}$ result in a purely positive potential across the whole domain of $z$ (thus the potential contains no bound states) there are combinations of $w_{i}$ which gives rise to potentials which are insufficiently deep and or wide to contain a bound state. The critical values of the potential parameters $w_{i}$ which guarantees the existence of a bound state are non-trivial. The threshold conditions are obtained by setting $E=0$ and $z=z_{1}$ and then calculating the zeros of equation~(\ref{eq:Wronskian}) as a function of $w_{i}$. The zeros of the Wronskian correspond to the values of $w_{i}$ for which a bound state emerges from the continuum. By determining the values of $w_{i}$ for which the first bound state emerges gives the critical values of $w_{i}$ which guarantees the existence of a bound state. %In figure~\ref{fig:critical} we plot the values of $w_{i}$ for which a new bound state emerges from the continuum.
It can be seen from figure~\ref{fig:critical} that the combination $w_{1}=15$,~$w_{2}=12$ and $w_{3}=1$ lies to the right of the critical values of $w_{i}$ which correspond to the emergence of the fourth bound state, therefore the said potential contains only three bound states.

\section{Conclusion}
It has been shown that a new class of hyperbolic asymmetric double-well
can be solved in terms of confluent Heun functions. An analytic procedure
for finding the eigenvalues via the calculation of the zeros of the Wronskian, constructed
from the two different Frobenius expansions about the two regular singularities
have been presented. The criterion for the potential to support bound states was discussed.
It is hoped that this model potential, with its easily found energy levels and multiple fitting parameters will serve as a useful tool in the study of phenomena whose behavior is described by asymmetric double-wells.
%anharmonic quantum potential related phenomena.

\section*{Acknowledgements}
This work was supported by URCO (17 N 1TAY12-1TAY13)

\end{document}